# APS Implementation over Vehicular Ad Hoc Networks

Soumen Kanrar
Vehere Interactive Pvt Ltd
Calcutta  India

**Abstract:** The real world scenario has changed from the wired connection to wireless connection. Over the year's software, development has responded to the increasing growth of wireless connectivity in developing network enabled software. The problem arises in the wireless domain due to random packet loss in transport layer and as well as in data link layer for the end- to-end connection. The  basic problem has been  considered in this work  is  to convert the real-world scenario of "Vehicular ad-hoc network" into a lab oriented problem by used the APS-system and study the result to achieve better performance in the wireless domain. The real-world physical problems map into analytical problem and simulate that analytic problem with respect to real-world scenario by Automated Position System (APS) for antenna mounted over the mobile node in 2 Dimension space. Here the methodology  quantifies the performance and the impact of the packet loss, delay, by the bit error rate and throughput with respect to the real- world scenario of VANET in the MAC layer, data link layer and transport layer. The result presents the Directional Antenna which is mounted over the vehicle gives less bit error in comparison to Isotropic and Discone antenna.

**Key words:** Automated Position System (APS), VANET, Bit error rate, Throughput, Antenna, Performance, Wireless connection.

## INTRODUCTION

Vehicular Ad Hoc Networks (VANET) differ from usual Ad Hoc Networks ((MANETs) in many different aspects. First VANETs consist of highly mobile nodes moving in the same or opposite directions. The nodes moving along the two different paths or curve. In shared wireless medium, blindly broadcasting packets may lead to frequent contention and congestion in transmission among neighboring nodes. This problem is some times referred to as the broadcast storm problem while multiple solution exit to alleviate the broadcast storm in the usual MANET environment only few. In this particular types of VANET problem, we considering the vehicle as the mobile node.

**Related work:** Gphan *et al*. (2007) discuss about the distance based schemes by using weighted p-persistence Broadcasting, slotted 1-persistence broadcasting and slotted p-persistence

Torrent *et al*. (2004) discusses about the broadcast performance scales in VANET.

Ni *et al*. (1999) various threshold-based techniques were proposed such as counter-based, distance based and location based

Hu *et al*. (2003) a directional antenna is used by Hu et al to mitigate broadcast redundancy and alleviate contention at the MAC layer.

Sze-Yao *et al*. (1999) discuss about the various mathematical model to reduce the redundancy, in the broadcast storm problem for mobile Ad hoc network.

Preston *et al*. (2005) authors considering the real world flight data and by using simulating check the validity.

Alphones and Saad (2009) authors worked on EAPS system by mounting isotropic antenna over the mobile node, see the performance of the message exchange in context client-server connection in wireless domain.

Stepanov and Rothermel (2007) author worked on the modeling of physical layer for MANET to developed realistic model "Energy consumption" for a device.

In present research the proposed schemes don't require a node to keep track of its neighbors.

**Proposed model:** In our proposed real time problem we considering in "the free space propagation model" a number of vehicles (mobile node) moving on the straight track and they maintain a minimum distance between them in 2D space Fig. 1. The road contains more than one track. We consider the vehicles move in a particular direction. The vehicles receive massage (alert massage-i.e. road bent, road block, bump ahead, road closed till side road like that) from the Broad caster ie Road side broadcaster unit. All communication is done through the fixed frequency. The road side broadcaster broadcast message and that will received by the vehicle (mobile node). We are going to the study the performance of system by using the antenna mounted over broadcaster (as isotropic ,directional and full cone)  and at the receiver mobile nodes isotropic antenna is mounted over that. The vehicles are moving node and the message broadcaster is a fixed base station. The proposed problem clearly concentrates into the wireless domain. The road side massage broadcaster broadcast the alert massage in a interval of time into the transmission region of the broadcaster and that massage is relayed between the neighbors mobile nodes (vehicle). So for the connection setup between the vehicle (mobile) nodes RTS/CTS required and for the forwarding of the alert message between the neighbors node fixed channel frequency is required. Packet collection and the problem of contention in that transmission region occurred.

Now for the experiment with the actual system, here we represent the experimental model of the system as the vehicle (mobile node) is the receiver represent by $R_x$ (with respect to antenna) and the road side broadcaster as the transmitter represent by $T_x$.

According to the Fig. 2, represent the locations A, D, C the positions of the vehicle in motions and B is the static position (broadcaster) at the different instant of time.

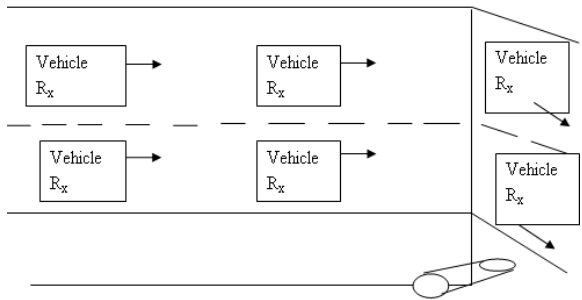

Fig. 1: Traffic alert system $T_x$

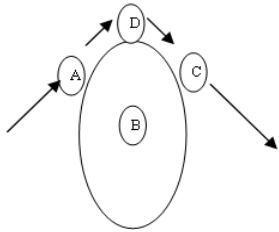

Fig. 2: Traffic Scenario of Vehicles

**Preliminaries:**

**Mobile Ad Hoc Network (MANET):** A MANET consists a set of mobile hosts that may communicate with one another from time to time. No base station is supported. Each host is equipped with a CSMA/CA (carrier sense multiple access with collision avoidance) transceiver. In such environment a host may communicate with another directly or indirectly.

**Redundant rebroadcast:** When a mobile host decides to rebroadcast message to its neighbors already have the message.

**Contention:** After a mobile host broadcast a message, if many of its neighbors decide to rebroadcast the message, these transmission ( which are all from nearby hosts) may severely contained with each other.

**Collision:** Because of deficiency of back off mechanism, The lack of RTS/CTS dialogue and the absence (CD) collision detection are more likely to occurred and create more damage.

**Directional antenna:** A directional antenna such as a parabolic antenna attempt to radiate most of its power in the direction of a known receiver.

**Isotropic antenna:** Isotropic antenna means is an antenna that transmits equally in all directions.

**Antenna gain:** It is power output, in a particular direction, compared to that produced in any direction by a perfect Omni directional antenna (isotropic antenna). Antenna gain is the measure in dB how much more power an antenna will radiate in a certain direction with respect to that which would be radiated by a reference antenna.

Relationship between antenna gain and effective area:

$$G = \frac{4\pi A_e}{\lambda^2} = \frac{4\pi f^2 A_e}{c^2}$$

G = Antenna gain
$A_e$ = Effective area
f = Carrier frequency
c = Speed of light
$\lambda$ = Carrier wavelength

**Cone antenna:** A conventional discone antenna is a cone antenna, this antenna typically include a feed structure that is within the cone. The cone antenna used in Wireless Local Area Network (WLAN).

**Azimuth angle:** The Azimuth angle, often denoted with a $\theta$ is the angle that the direct transmission makes with respect to a given reference angle (often the angle of the target receiver) when looking down on the antenna from above.

**Elevation angel:** The elevation angle is the angle that the transmission direction makes with the ground elevation angle is denoted with a $\phi$.

**Free space model:** The free space propagation model assumes the ideal propagation condition that there is only one line of-sight path between the transmitter and receiver.

Friis presented the following equation to calculate the received signal power in the free space at distance d from the transmitter:

$$P_r(d) = \frac{P_t G_t G_r \lambda^2}{(4\pi)^2 d^2 L}$$

Here,
$P_t$ = The transmitted signal power
$G_t, G_r$ = The antenna gains of the transmitter and receiver respectively
L = ($1 \leq L$) is system loss
$\lambda$ = The wavelength

**Simulation model:** OPNET was used to build the simulation model. All the operations are done by using OPNET kernel procedures. This is Baseline simulation. The Antenna gain pipeline stage are modified at the transmitter and receiver according to the model requirement. The role of the pipeline stage is to compute the antenna gain, throughput and Bit error rate, for the different position of the mobile node.

The Enhanced Antenna Position System (EAPS) was used to improve the performance gain. In general EAPS used two modes:

- Fixed to object: The fixed to object mode is describing the movement of an antenna that mounted over the mobile node and moves with the mobile node
- Locked to target: In this mode the antenna always points at a specific target

In both the modes, rotation angle parameter are used so the antenna rotate about its pointing axis (

**Scenario description:** The system consists of a fixed transmitter and mobile node receiver with mobile Jammer moving in a trajectory in the area around 8000×4000 m. The receiver is model by using random waypoint mobility i.e., the receiver moves randomly. The jammer node is used to create radio noise. Here the transmitter and receiver modules use different channel for connection setup by using the control packet and for message broadcasting and for message forwarding by using the data packet. The jammer always change its position around the receiver by travels its path. The simulation is run for 12 min and the size of the packet send by the transmitter is 1024 bits. The transmitter broadcast each packet in 1 min. The transmission power at the transmitter is 20 watts. The receiver node moves with velocity of 10 m sec$^{-1}$. Figure 3 represent the receiver antenna attributes.
**Node model:**

**Transmitter:** The transmitter composed of three modules via Fig. 4:

- Simple source i.e., packet generator
- Antenna target tracker module. This module tracks the antenna location
- Radio transmitter module. This module transmits the packets on a radio channel

**Receiver:** The receiver composed of three module via Fig. 5:
- Antenna module
- Radio receiver module: This module defined the gain which will be adjust at run time
- Sink processor module: This module store the received packets

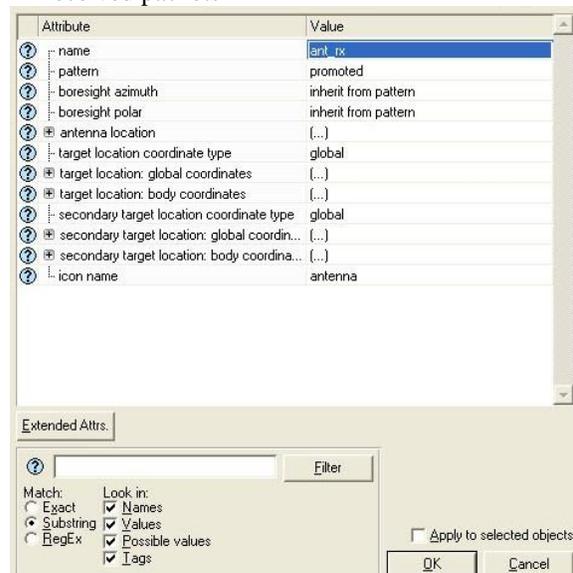

Fig. 3: Receiver antenna setting

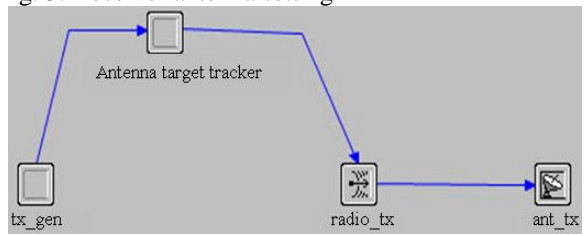

Fig. 4: Transmitter Model

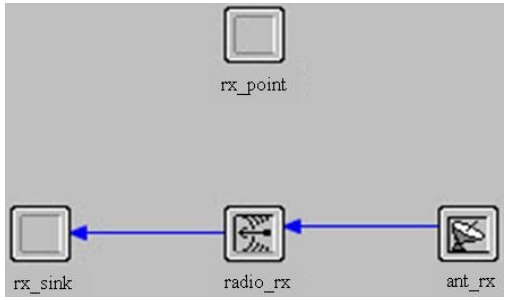

Fig. 5: Receiver Model

Table 1: simulation parameters

| | |
|---|---|
| Transmission power | 20 w |
| Simulation time | 12 min |
| Movement area | 8×4 km |
| Node movement speed | 10 m sec$^{-1}$ |
| Modulation | BPSK |
| Packet size | 1024 bits |
| Packet interval time (broadcast) | 1.0 m |

**Performance analysis:** For the performance of the system we study the Bit error rate and the throughput for the different types of antenna. Such as isotropic, directional, full cone antenna mounted over the transmitter and the isotropic over the receiver. We put the jammer in a trajectory that artificially improve the situation as real world, the jammer and the transmitter produce packet independently and asynchronously. In the Fig. 6 when the vehicles approach toward the transmitter, the received signal strength is increases. when the vehicles at the point A according to Fig. 2.

Since |AB|<|AD| the received signal strength decrease and again increase at the point C according to Fig. 2, |BD|>|BC| and again decrease. According to Fig. 6, at the time 5.5 and 7.5 min the receiver by using isotropic get maximum signal strength. Where the receiver with directional and full cone antenna get maximum signal strength at the 5.5 min.

Bit error rate concern about the rejection of packet, when it received with error. Where in the case of throughput, how many packets received properly with respect to the packet transmitted from the transmitter in presence of the jammer. The jammer artificially generate noise packet according to the Fig. 7 and 8, the simulation done by packet to packet basis. In the case of received power, we concern about the signal strength at the receiver side in compare to the transmitted signal power Fig. 6. Where the jammer artificially generate noise as the real life scenario.

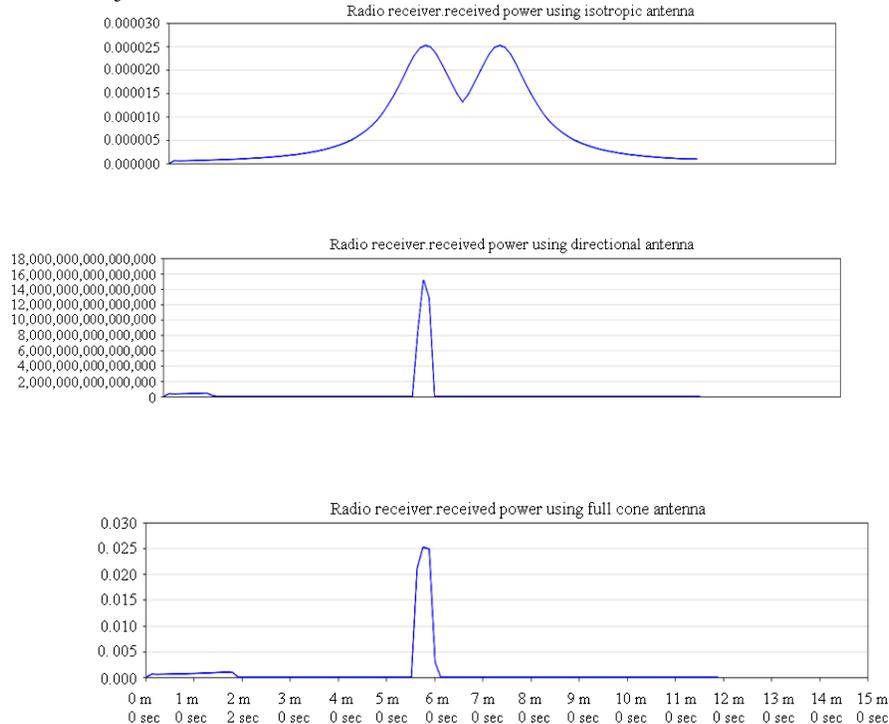

Fig. 6: Radio receiver power

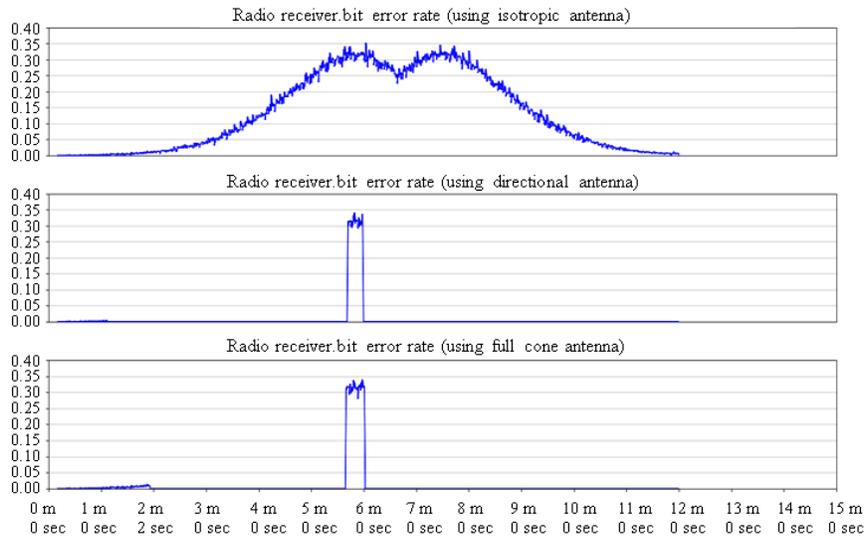

Fig. 7: Bit error rate with no antenna tracker

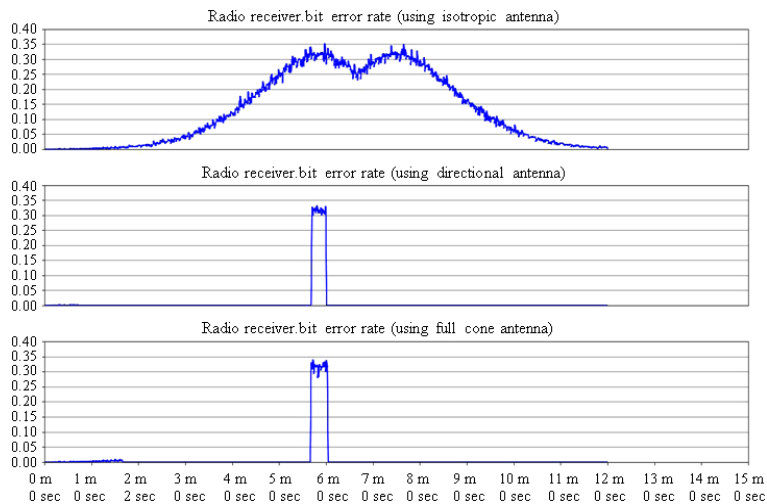

Fig. 8: Bit error rate with antenna tracker

Figure 7 and 8 represent the bit error rate without antenna tracker and with antenna tracker. Bit error rate initially zero at the receiver node as the distance between the jammer node and receiver node is large. After 5.5 min the direction vector between the jammer antenna and the receiver antenna was in the line with the direction of gain for receiver antenna. When the receiver node get maximum bit error during the interval 5.5-6.0 min via-Fig. 7 and 8 for different types of antenna which are mounted over the receiver node.

Figure 9 and 10 represent the throughput of the system with antenna tracker and without antenna tracker. Clearly it seen from the Figure 10 during the time interval 5.5-6.0 min the throughput is approaches to zero where the Bit error rate is maximum.

In the Fig. 9 and 10, we have seen that initially the throughput is maximum for isotropic antenna and it will decrease to zero with the increase of time. For the other types of antenna like directional or cone antenna via Fig. 10 the throughput of the system goes up and down during the interval 0.0-1.0 min. For the directional antenna via Fig. 9 throughput all most approaches maximum at 1.0 min and for the full cone antenna the system throughput approaches maximum at the 2.0 min and gives zero throughput at 6.0 min where the bit error rate is maximum via Fig. 7 and 8 0.33 errors/bits.

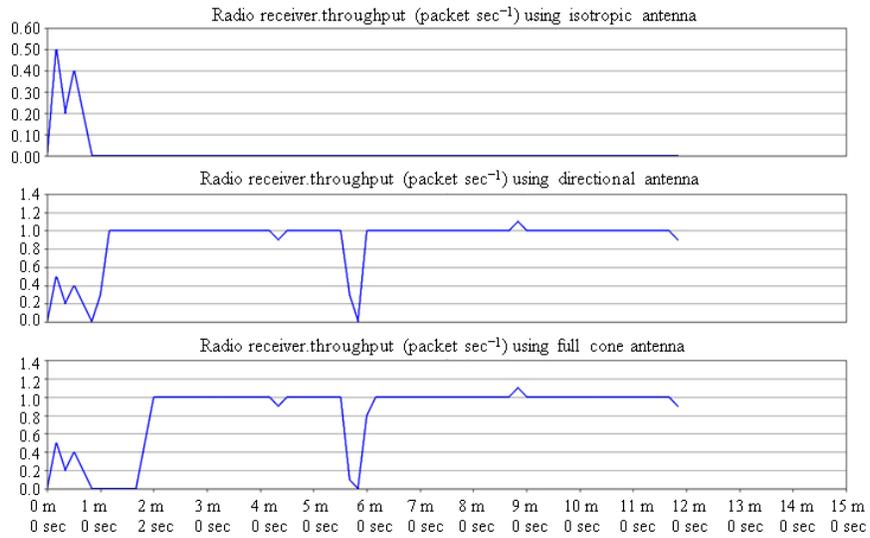

Fig. 9: Throughput with no antenna tracker

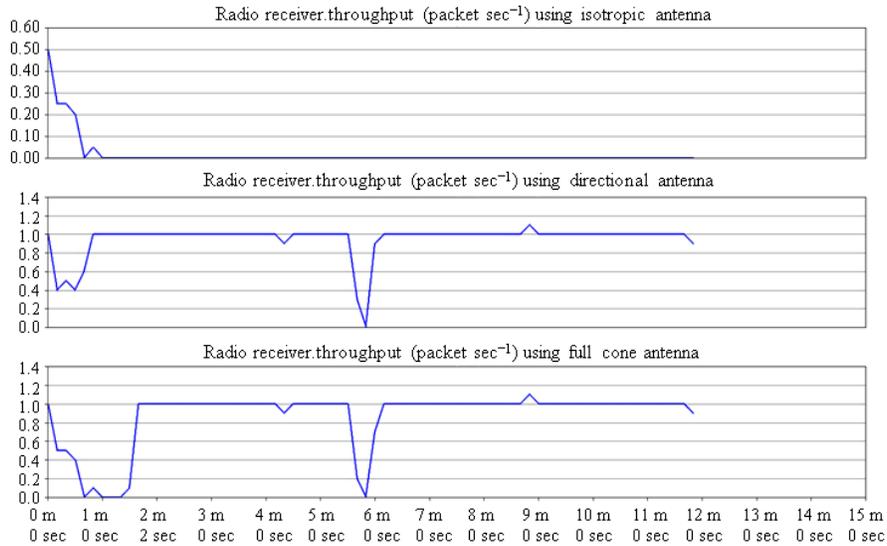

Fig. 10: Throughput with antenna tracker

In Fig. 10 we have seen the throughput without the antenna tracker of the system. Initially the throughput of the system is maximum for the isotropic antenna and gradually decrease approach to zero at 1.5 min. The system gets maximum throughput at 1.0 and 1.5 min for other type of antenna used at the receiver node. The system get zero throughput at 6.0 min where the Bit error is maximum.

**RESULT and DISCUSSION:**

In this work we have used OPNET 14.5 Modeler. All the operations have been done by using OPNET kernel procedures. We have used the default radio pipeline of OPNET figure -11

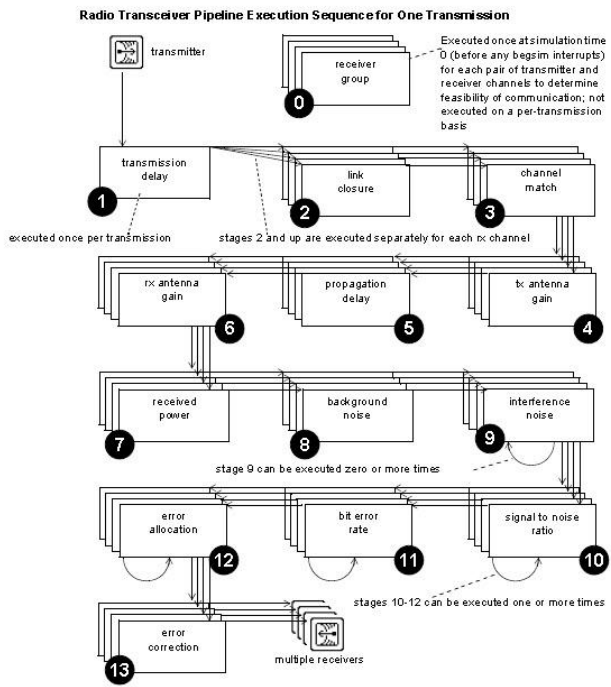

Figure -11 Opnet Pipeline Stages

From the results we observe that the Directional Antenna which is mounted over the vehicle gives less bit error in comparison to Isotropic and Discone antenna,

## CONCLUSION

In this work, extended open for the modeling of MANET applications in the city scenarios. This work applied more realistic mobility and wireless transmission implemented in the vehicular ad hoc network. This work shown that how the packet loss, i.e., bit error rate and that effect on the throughput of the system performance in real scenario. Obviously, more realistic models come at a cost of increase complexity of the model as well as in the protocol development. So better traffic modeling and the transport planning is required for the dynamics model.
.